\documentclass[a4paper]{jpconf}
\usepackage{hyperref}
\usepackage{epsfig}

\begin{document}

\title{Nonlinear hysteretic behavior of a confined sliding layer}

\author{N. Manini$^{1,2}$,
        G.E. Santoro$^{2,3}$, E. Tosatti$^{2,3}$,
	and  A. Vanossi$^4$}
\address{
$^1$Dipart. di Fisica and CNR-INFM, Universit\`a di Milano, Via Celoria 16, 20133 Milano, Italy
}
\address{
$^2$International School for Advanced Studies (SISSA)
and CNR-INFM Democritos National Simulation Center, Via Beirut 2-4, I-34014 Trieste, Italy
}
\address{
$^3$International Centre for Theoretical Physics (ICTP), P.O.Box 586, I-34014 Trieste, Italy
}
\address{
$^4$CNR-INFM National Research Center S3 and Department of Physics,
University of Modena and Reggio Emilia, Via Campi 213/A, 41100 Modena, Italy
}

\begin{abstract}
A nonlinear model representing the tribological problem of a thin solid
lubricant layer between two sliding periodic surfaces is used to analyze
the phenomenon of hysteresis at pinning/depinning around a moving state
rather than around a statically pinned state.
The cycling of an external driving force $F_{\rm ext}$ is used as a simple
means to destroy and then to recover the dynamically pinned state
previously discovered for the lubricant center-of-mass velocity.
De-pinning to a quasi-freely sliding state occurs either directly, with a
single jump, or through a sequence of discontinuous transitions.
The intermediate sliding steps are reminiscent of phase-locked states and
stick-slip motion in static friction, and can be interpreted in terms of
the appearance of travelling density defects in an otherwise regular
arrangement of kinks.
Re-pinning occurs more smoothly, through the successive disappearance
of different travelling defects.
The resulting bistability and multistability regions may also be explored
by varying mechanical parameters other than $F_{\rm ext}$, e.g.\ the
sliding velocity or the corrugation amplitude of the sliders.
\end{abstract}



\section{Introduction}

Nonlinear systems driven far from equilibrium exhibit a very rich
variety of complex spatial and temporal behaviors \cite{Kapitaniak99}.
In particular, in the emerging field of nanoscale science and technology,
understanding the nonequilibrium dynamics of systems with many degrees of
freedom which are pinned in some periodic potential, as is commonly the
case in solid-state physics, is often becoming an issue.
Friction belongs to this category too, because the microscopic corrugation
of the mating surfaces may interlock \cite{Perssonbook, Rubinstein04}.
Simple phenomenological models are important, as they often give not only
qualitative understanding of experimental findings, but also fair
quantitative agreement with
%
nanoscale tribology data, 
and with realistic simulations of sliding phenomena \cite{Vanossi_review}.
In this line of simplified approaches, studies are typically restricted to
describing microscopic dynamics in one (1D) or two (2D) spatial dimensions.
The substrates defining the moving interface are modelled in a simplified
way as purely rigid surfaces or as one- or
two-dimensional arrays of particles interacting through simple (e.g.,
harmonic) potentials.
Despite such a crude level of description, this class of approaches
frequently reveals the ability of modelling the main features of the
complex microscopic dynamics, ranging from regular to chaotic motion
\cite{Braunbook,Rozman96,Zaloj98}.

One of the pervasive concepts of modern tribology -- with a wide area of
relevant practical applications as well as fundamental theoretical issues
-- is the idea of free sliding connected with {\it incommensurability}.
When two crystalline workpieces with incommensurate or misaligned lattices
are brought into contact, the minimal force required to achieve sliding,
i.e.\ the static friction, should vanish, at least provided the two
crystals are stiff enough.
In such a geometrical configuration, the lattice mismatch can prevent
interlocking of the two periodic corrugations
and the resulting collective stick-slip motion of the interface atoms, with
a consequent dramatically reduced frictional force.
Experimental observation of this sort of {\it superlubric} and anisotropic
regime of motion has been reported recently \cite{Dienwiebel04,Salmeron}.
The paradigm of frictionless sliding is realized naturally by the 1D
Frenkel-Kontorova (FK) model (see Ref.~\cite{Braunbook} and references
therein).
However the physical contact between two solids is generally mediated
by so-called ``third bodies'', and the role of incommensurability has been
recently extended \cite{Braun05} in the framework of a driven 1D model
inspired by the tribological problem of two sliding interfaces with a thin
solid lubricant layer in between.
The frictional interface is thus characterized by {\it three} inherent
length scales along the sliding direction: the periods of the bottom and
top substrates, and the period of the embedded solid lubricant structure.
In particular, in the presence of a uniform external driving velocity, the
interplay of these incommensurate length scales can give rise to intriguing
dynamical phase locking phenomena and surprising velocity quantization
effects \cite{Vanossi06,Santoro06}.

Previous numerical and theoretical studies of this confined tribological
model \cite{Vanossi06,Santoro06,Manini07extended} discovered a quantization
of the lubricant center-of-mass (CM) relative velocity and found it to be
related to the pinning of topological density excitations (kinks) to the
substrate of closest periodicity.
More recent work \cite{Vanossi07PRL} highlighted a strict analogy of these
dynamical pinning phenomena to the ordinary commensurate pinning of {\it
static} friction \cite{Braun97,Ariyasu87}.
The proposed mapping between this dynamical pinning and that of static
friction was explored numerically by analyzing the effect of an additional
external driving force $F_{\rm ext}$, equal for all lubricant particles.
Dynamical pinning is signified by the lubricant CM relative velocity
remaining robustly locked to the quantized plateau value (a value strictly
and analytically determined by spatial periodicity ratios alone)
up to a critical force threshold, above which quantization is destroyed.

It was also found that as long as inertial effects are non negligible
compared to dissipative forces ({\em underdamped} regime of motion), the
adiabatic variation (increase and decrease) of the external driving force
gives rise to a large hysteresis loop in the $v_{\rm cm}$ -- $F_{\rm ext}$
characteristics, not unlike depinning in static friction
\cite{Braunbook,Braun97}.
The present paper focuses precisely on the hysteretic behavior around a
dynamical quantized steady state that this system exhibits, and discuss
similarities and differences between such a dynamical locking and the more
usual static pinning.
By exploiting configurations where the dynamics of individual kinks is easy
to monitor visually, the mechanism of hysteresis will be clarified.
Given the practical difficulty of an experimental setup where an equal
driving force is applied to each lubricant particle on the fly,
the $F_{\rm ext}$ term is may be seen more as a useful mathematical device
rather than a realistic suggestion for future measurements aimed at
studying dynamical depinning.
On the other hand, we will bring concrete examples of the hysteretical
destruction and recovery of the CM velocity plateau by means of parameters
other than $F_{\rm ext}$ being cycled.
The cycling of the substrate sliding velocity or of the applied load
sketch practical possibilities to address the dynamical hysteresis in
experimental tribological investigations.

\section{Confined lubricant model: numerical simulations}

\begin{figure}
\centerline{\epsfig{file=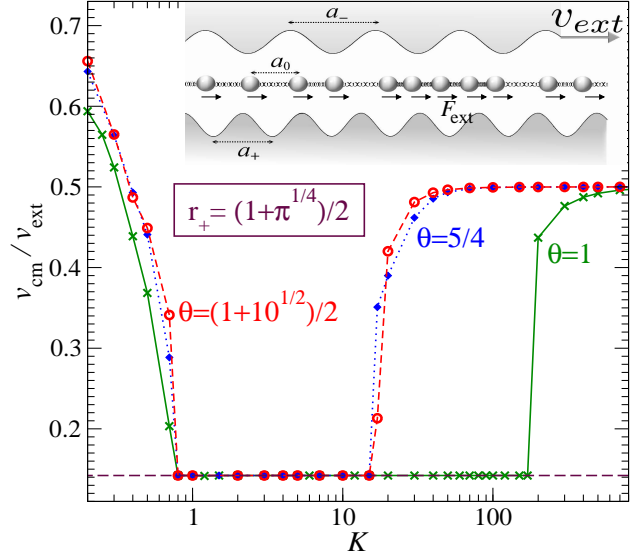,width=8.2cm,angle=0,clip=}}
\caption{\label{velcm_model}
(Color online) Normalized velocity of the center of mass, $v_{\rm
cm}/v_{\rm ext}$, as a function of the chain stiffness $K F_+/a_+$, for
$v_{\rm ext}=0.1 \,(F_+a_+/m)^{1/2}$, $\gamma=0.1 \,(F_+m/a_+)^{1/2}$, and
$r_+=(1+\pi^{1/4})/2$.
Crosses: one-to-one kink coverage $\theta=1$ ($r_-\simeq 7.036$); diamonds:
commensurate kink coverage $\theta=5/4$ ($r_-\simeq 8.795$); circles:
incommensurate kink coverage $\theta=(1+10^{1/2})/3$ ($r_-\simeq 9.762$).
Dashed line: the quantized-plateau velocity ratio of Eq.~(\ref{vplateau}).
Note the logarithmic scale in the abscissa.
Inset: a sketch of the driven 3-lengthscale confined model.
}
\end{figure}

We will work with the one-dimensional generalization of the standard FK
model introduced in Ref.~\cite{Vanossi06,Santoro06}, consisting of two
rigid sinusoidal substrates, of spatial periodicity $a_{+}$ and $a_{-}$,
and a chain of harmonically interacting particles, of equilibrium length
$a_0$, mimicking the sandwiched lubricant layer, as sketched in the inset
of Fig.~\ref{velcm_model}.\footnote{%
The harmonicity of interactions within the lubricant chain is merely a
simplifying assumption, since test simulations with anharmonic
inter-particle potentials (e.g.\ Morse and Lennard-Jones) also reveal the
ubiquity of the observed phenomenology.
}
The two substrates move at a constant relative velocity $v_{\rm ext}=v_- -
v_+$.
In particular, we select the reference frame where $v_+ = 0$ and $v_- =
v_{\rm ext}$.
The equation of motion of the $i$-th lubricant particle is:
\begin{eqnarray} \label{eqmotion:eqn}
m\ddot{x}_i = -\frac{1}{2} \left[ F_+ \sin{\frac{2\pi}{a_+}
x_i} + F_- \sin{\frac{2\pi}{a_-} (x_i-v_{\rm ext}t)}\right] \nonumber \\
+ K (x_{i+1}+x_{i-1}-2x_i) -2\gamma (\dot{x}_i - v_{\rm w})+
F_{\rm ext} \;,
\end{eqnarray}
where $m$ is its mass. $F_{\pm}$ are the amplitudes of the forces
due to the sinusoidal corrugation of the substrates.
By default, we set $F_-/F_+=1$ as the least biased choice, but we will
explore the effect of modifying $F_-$ in Sect.~\ref{results} below.
$K$ is the chain spring constant defining the harmonic nearest-neighbor
interparticle interaction.
The penultimate damping term in Eq.~(\ref{eqmotion:eqn}) originates from
two symmetric frictional contributions adding as follows:
$-\gamma\,(\dot{x}_i - v_+) -\gamma\,(\dot{x}_i - v_-) =
-2\gamma\,(\dot{x}_i - \frac 12 v_{\rm ext})$, where $\gamma$ is a viscous
friction coefficient accounting phenomenologically for degrees of freedom
inherent in the real physical system (such as substrate phonons, electronic
excitations, etc.) which are not explicitly included in the model; this
fixes the reference speed of the the dissipative term: $v_{\rm w}=\frac 12
v_{\rm ext}$ \footnote{%
We choose this value of the velocity $v_{\rm w}=\frac 12 v_{\rm ext}$ to
which dissipation refers as the least biased option.
Different choices, equivalent to choosing different $\gamma_+$ and
$\gamma_-$ dissipation coefficients to the two substrates, would at most
change the quantitative details of the velocity-plateau boundaries, but not
the qualitative nature of results.
}.
In order to probe the strength of quantization, and eventually address
hysteresis, an additional constant force $F_{\rm ext}$ is applied to all
chain particles and varied up and down adiabatically.
The infinite chain size is managed -- in the general incommensurate
case -- by means of periodic boundary conditions (PBC) and finite-size
scaling \cite{Manini07extended}. 
We set overall $a_+=1$, $m=1$, and $F_+=1$ as basic dimensionless units,
and express implicitly all mechanical quantities in terms of natural model
units obtained as combinations of these three basic units
\cite{Manini07extended}.

As previously found \cite{Vanossi06,Santoro06,Manini07extended,Vanossi07PRL}
the detailed behavior of the driven system in Eq.~(\ref{eqmotion:eqn})
depends crucially on the relative (in)commensurability of the substrates
and the chain. The relevant length ratios are defined by
$r_{\pm}=a_{\pm}/a_0$; we assume $r_- >
\min(r_+,r_+^{-1})$, whereby the $(+)$ substrate has the closest periodicity
to the lubricant, the $(-)$ slider the furthest.
Under rather general dynamical conditions, the lubricant slides with a
quantized mean velocity $v_{\rm plateau}$ relative to the $(+)$ substrate.
The plateau phenomenon was explained by the static pinning of the
topological solitons (kinks) that the embedded chain forms with the $(+)$
substrate, to the $(-)$ slider \cite{Vanossi06,Manini07extended}.
Specifically, the quantized-plateau lubricant velocity ratio
\begin{equation}\label{vplateau}
\frac{v_{\rm cm}}{v_{\rm ext}} =
\frac{v_{\rm plateau}}{v_{\rm ext}} \equiv
1 - \frac 1{r_+}\,,
\end{equation}
is strictly a function of the lubricant coverage $r_+$ of the $(+)$
substrate \cite{Vanossi06}, i.e.\ of the absolute density $(r_+ -1)/a_+$ of
kinks. For antikinks, $r_+<1$, this density is negative, and so is $v_{\rm
plateau}$-- namely the lubricant slides {\em backwards} \cite{Vanossi06}.
Although the quantized-plateau velocity depends uniquely on $r_+$, the
plateau dynamical stability and extension depend crucially on the kink
coverage
\begin{equation}\label{theta}
\theta=a_-\,\frac{r_+ -1}{a_+}
=r_- \left(1-\frac 1{r_+}\right)
\end{equation}
of the $(-)$ substrate (for antikinks, $\theta<0$).
Concretely, as a function, e.g.\ of the spring stiffness $K$, the quantized
plateau is very prominent in a range of $K$ of the order unity but weakens
and eventually terminates for stiffer chains (larger $K$ values),
Fig.~\ref{velcm_model}.
The plateau destabilization is complete for a general irrational $\theta$,
while, under suitable conditions detailed below, the plateau can survive up
to indefinitely large $K$ for commensurate kink coverage (rational
$\theta$).
The quantized velocity plateau is finally particularly robust for perfect
one-to-one matching of the soliton and the (-) slider periodicities,
$\theta=1$ \cite{Vanossi07PRL}.
To illustrate these three typical cases, we consider
$r_+=(1+\pi^{1/4})/2\simeq 1.166$ and the three values $r_-\equiv \theta
\,(1-r_+^{-1})^{-1} \simeq 7.036$, $8.795$, and $9.762$, corresponding to the
values $\theta=1$, $\theta=5/4=1.250$, and $\theta=(1+10^{1/2})/3\simeq
1.387$ respectively.
The choice of $r_+$ near unity, is especially advantageous compared to
values like the golden mean $(1+\sqrt{5})/2\simeq 1.618$ often used,
because it gives rise to well-separated individual kinks, which allow a
more transparent analysis of the dynamics.
Many qualitative features discussed for the specific ratios $r_\pm$
considered here are in fact also found for general values of $r_\pm$, and
thus this specific choice of length ratios should not be considered
especially restrictive, as long as a correct distinction of different
commensuration property of $\theta$, Eq.~(\ref{theta}), is made.

The equations of motion (\ref{eqmotion:eqn}) are integrated using a
standard fourth-order Runge-Kutta algorithm. The system is initialized with
the chain particles placed at rest at uniform separation $a_0$, and the top
substrate is made slide at the imposed constant velocity $v_-=v_{\rm ext}$.
For $F_{\rm ext}=0$ and a wide range of model parameters, after an initial
transient the system reaches a steady state, where all dynamical quantities
other than particle positions fluctuate but show no systematic drift.
For wide ranges of parameters, exemplified in Fig.~\ref{velcm_model} by
the spring stiffness $K$, the lubricant reaches the expected plateau state
of normalized time-averaged velocity $v_{\rm plateau}/v_{\rm ext}\simeq
0.142$, Eq.~(\ref{vplateau}), the same for the three geometries introduced
above.

Adiabatic upward and downward variation of the external force $F_{\rm
ext}$ is realized by changing $F_{\rm ext}$ in small steps and letting the
system evolve at each step for a time long-enough for all transient
stresses to relax. This allows us to gauge the robustness of the plateau state
as a function of the system parameters, e.g.\ of $K$.
In order to determine the critical values of $F_{\rm ext}$, where the
plateau is abandoned and retrieved, and in particular to do that with great
accuracy and a reasonably small number of separate simulations, we first
increment $F_{\rm ext}$ in steps of $0.01\,F_+$, and then reduce the step
width using a bisection scheme around the critical force.

\section{Results}\label{results}

\begin{figure}
\centerline{\epsfig{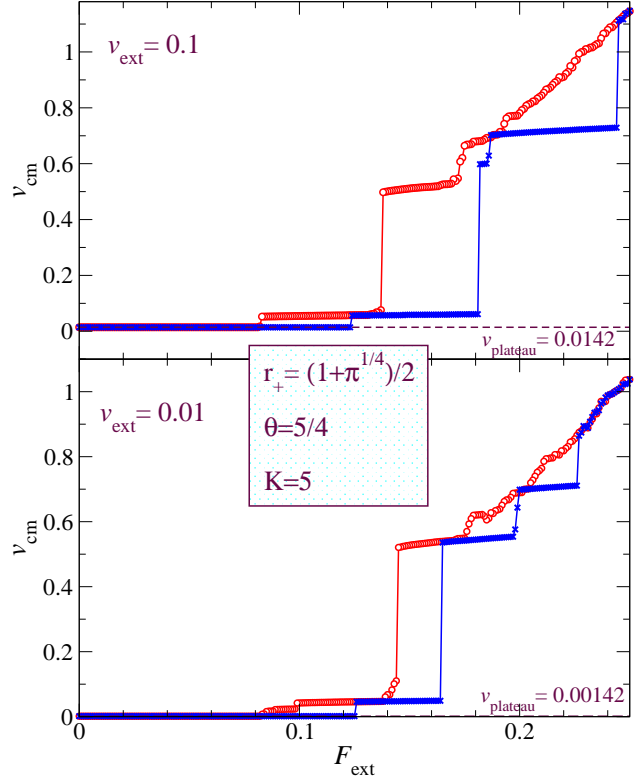}}
\caption{\label{loop_hyst}
(Color online)
Hysteresis in the $v_{\rm cm}-F_{\rm ext}$ characteristics for a confined
chain of intermediate spring stiffness ($K=5 F_+/a_+$), and length ratios
$r_+=(1+\pi^{1/4})/2$, $\theta=5/4$ ($r_-\simeq 8.795$).
The behavior is shown for fast ($v_{\rm ext}=0.1$, upper panel) and slow
($v_{\rm ext}=0.01$, lower panel) drive.
Adiabatic increase and decrease of $F_{\rm ext}$ (in steps of $10^{-3} F_+$)
are denoted by crosses and circles, respectively.
Characteristic hysteretic multi-step features appear.
Here $\gamma=0.1 (F_+m/a_+)^{1/2}$, and a chain of $N=387$ lubricant
particles is simulated.
}
\end{figure}

For concreteness, we begin with the specific example $\theta=5/4$, and pick
an intermediate value of the chain stiffness $K = 5\,F_+/a_+$, common to
all plateaus of Fig.~\ref{velcm_model}.
We start investigating the plateau destruction/recovery induced by varying
the external force $F_{\rm ext}$ through a sequence of adiabatic increases
and decreases \cite{Vanossi07Hyst}.
The resulting CM velocity is displayed in Fig.~\ref{loop_hyst}
for two different external driving velocities $v_{\rm ext}$.
A clear hysteretic loop emerges, with qualitatively similar features for high
(upper panel) and low (lower panel) values of $v_{\rm ext}$.
Interestingly, and somewhat unexpectedly, the hysteretic regions are
systematically broader for larger sliding velocities $v_{\rm ext}$.
We will return to this point later on.

The exact plateau state implies a kind of {\em dynamical
incompressibility}, namely identically null response to perturbations or
fluctuations trying to deflect the CM velocity away from its quantized
value.
Indeed, as long as $F_{\rm ext}$ remains below a critical threshold
$F_c^{+\uparrow}$, it does perturb each individual single-particle motion,
but has no effect whatsoever on $v_{\rm cm}$, which remains exactly pinned
to the quantized value, as is indeed expected of an incompressible state.
This behavior contrasts with all observed non-plateau sliding states, where
$v_{\rm cm}$ increases monotonically with $F_{\rm ext}$.
This plateau state is reminiscent of the pinned state of static friction,
where a minimum force (the static friction force) is required to initiate
the motion.
Except that here in the starting ``pinned'' plateau state the lubricant
chain particles are {\em moving} relative to both substrates.
The sudden upward jump of $v_{\rm cm}$ taking place at $F_{\rm ext}
=F_c^{+\uparrow}$ can thus be termed a {\em dynamical depinning}.
The depinning transition line $F_c^{+\uparrow}$, appears as a
``first-order'' line, with a finite jump $\Delta v$ in the average $v_{\rm
cm}$ and a clear hysteretic behavior: as $F_{\rm ext}$ is reduced back, the
depinned state survives below $F_c^{+\uparrow}$ down to a significantly
smaller $F_c^{+\downarrow}$, where perfectly quantized plateau sliding is
retrieved, as illustrated in Fig.~\ref{loop_hyst}.
Several hysteretic loops are in fact observed in Fig.~\ref{loop_hyst}:
a qualitatively similar multi-step behavior appears also for $\theta=1$ and
$\theta=(1+10^{1/2})/3$.
We shall return below to the nature of these steps.
The large-$F_{\rm ext}$ quasi-free sliding regime is characterized by
$v_{\rm cm}$ increasing continuously, roughly proportionally to $F_{\rm
ext}/\gamma$, and superposed to this general translational motion, by
chaotic single-particle movements, contrasted to the periodic ($\theta=1$,
$5/4$) or quasi-periodic ($\theta=(1+10^{1/2})/3$) individual-particle
oscillations in the plateau state.

\begin{figure}
\centerline{\epsfig{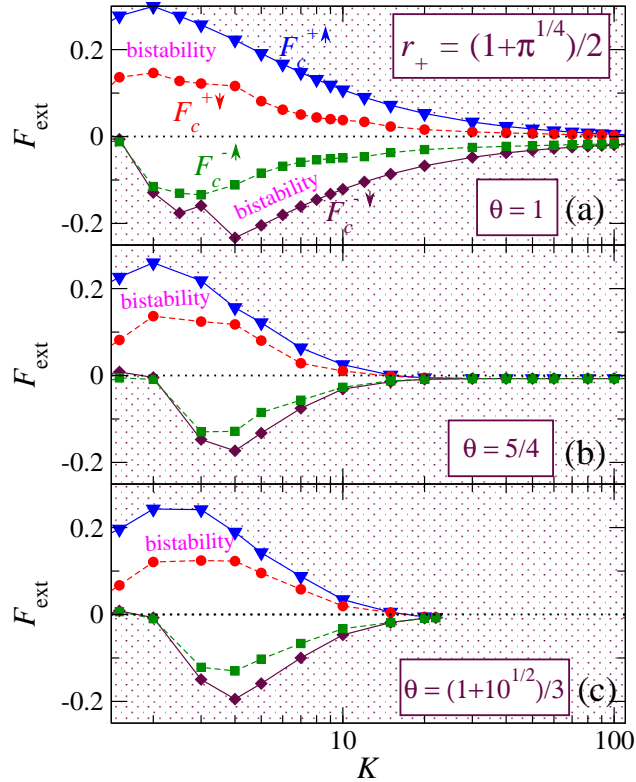}}
\caption{\label{criticalF}
(Color online)
The $(K,F_{\rm ext})$ phase diagram illustrating the unpinning-repinning
transitions for $r_+=(1+\pi^{1/4})/2$, $\gamma=0.1$, $v_{\rm ext}=0.1$, and
for
({\bf a}) one-to-one kink coverage $\theta=1$;
({\bf b}) commensurate kink coverage $\theta=5/4$;
({\bf c}) incommensurate kink coverage $\theta=(1+10^{1/2})/3$.
The white areas have perfect plateau dynamics, the dotted region indicates
quasi-free sliding.
Simulations done with $N=387$ particles for ({\bf a}) and ({\bf b}) and
with $N=781$ particles for ({\bf c}).
}
\end{figure}

The values of $F_c^{+\uparrow}$, $F_c^{+\downarrow}$, and $\Delta v$ are
nontrivial functions of the parameters.
Specifically, Fig.~\ref{criticalF} reports the $K$ dependency of these
critical forces in the three considered cases.
The values of the critical forces are remarkably similar for $K < 4$,
while important differences are observed as the springs become stiffer.
In particular, for unity coverage ($\theta=1$) the plateau is very stable
and extends to very large $K$, as expected in a fully commensurate case,
see Fig.~\ref{criticalF}(a).
In contrast, for noninteger $\theta$ the plateau becomes more fragile for
large $K$.
For commensurate $\theta=5/4$ the plateau width decreases with some fast
power law of $K$, and becomes numerically difficult to detect beyond
$K\simeq 60$.
For incommensurate $\theta=(1+10^{1/2})/3$ instead, the plateau shrinks and
disappears at finite $K=K_{\rm Aubry}^{\rm dyn}\simeq 24$: no sign of a
quantized plateau is detectable, e.g., for $K=25$.
This unequal behavior for commensurate/incommensurate coverage $\theta$ is
understood in terms of the mapping of the dynamical sliding model to the
static FK model, which was established in Ref.~\cite{Vanossi07PRL}.
The hysteretic depinning transition is observed through a significantly
wide $K$-range in all three cases, but the depinning mechanism differs in
some important detail.

\subsection{Fully commensurate $\theta=1$}

As illustrated in Fig.~\ref{criticalF}, for $\theta=1$ the plateau extends
to very large $K$, in a range of $F_{\rm ext}$ of decreasing width $\propto
K^{-1}$.
$K^{-1}$ describes precisely the asymptotic decrease of the sinusoidal
interparticle distance modulation, residual after solitons overlap one
another in the large $K$ limit.
For very large $K$, outside the right end of Fig.~\ref{criticalF}(a), the
asymptotic values of this $F_c^{+\uparrow}$ curve lie entirely in the
negative-$F_{\rm ext}$ domain.
The explanation is that it takes a negative external force to compensate
the positive average dissipative ``wind'' force $F_{\rm w}=-2\gamma (v_{\rm
cm} - v_{\rm w})$ acting on each lubricant particle.
On the plateau state this wind force amounts to
\begin{equation}\label{Fw}
F_{\rm w}=
-2 \gamma\,(v_{\rm plateau}-v_{\rm w})
=
\frac {2-r_+}{r_+}\,\gamma\,v_{\rm ext}
\,.
\end{equation}
In the absence of the external driving $F_{\rm ext}$, the wind force alone
is sufficient to disrupt the plateau at large $K$, where it is more
fragile, as seen at the large-$K$ side of Fig.~\ref{velcm_model}.
However, once $F_{\rm w}$ is compensated away, the  $\theta$ =1
quantized plateau extends to indefinitely large $K$.

The next result concerns hysteresis, still at $\theta$ =1.
Depinning is discontinuous and hysteretical as exemplified in
Fig.~\ref{loop_hyst}, but only up to a large but finite critical stiffness
$K=K_*\simeq 330$.
Near $K_*$ the bistability range $F_c^{+\uparrow} -F_c^{+\downarrow}$
closes up with a power law $F_c^{+\uparrow} -F_c^{+\downarrow} =
B\,(K_*-K)^\alpha$, not unlike what was observed in previous work for the
golden-mean ratio Ref.~\cite{Vanossi07PRL}.
Above the critical stiffness, for $K\geq K_*$, $F_c^{+\uparrow}\equiv
F_c^{+\downarrow}$, the depinning transition is continuous and
characterized by what appears to be a mean-field power law $v_{\rm
cm}-v_{\rm plateau}\propto (F_{\rm ext}- F_c^{+\downarrow})^{1/2}$.
For $K$ approaching $K_*$ from below, the plateau is abandoned through
different mechanisms depending on the model parameters.
In Ref.~\cite{Ariyasu87} it was found that re-pinning in the
continuous sine-Gordon model proceeds first through a series of
``cavity-mode'' states, and then a series of kink-antikink wave train
states, and a similar scenario is exhibited also by the discrete FK chain
\cite{Braun97}. 
We find that analogous phenomena occur here for the repinning to the
dynamical plateau, with defects in the kink lattice taking the place of the
kink-antikink pairs of the single-chain FK model.

\begin{figure}
\centerline{\hfill(a)
\epsfig{file=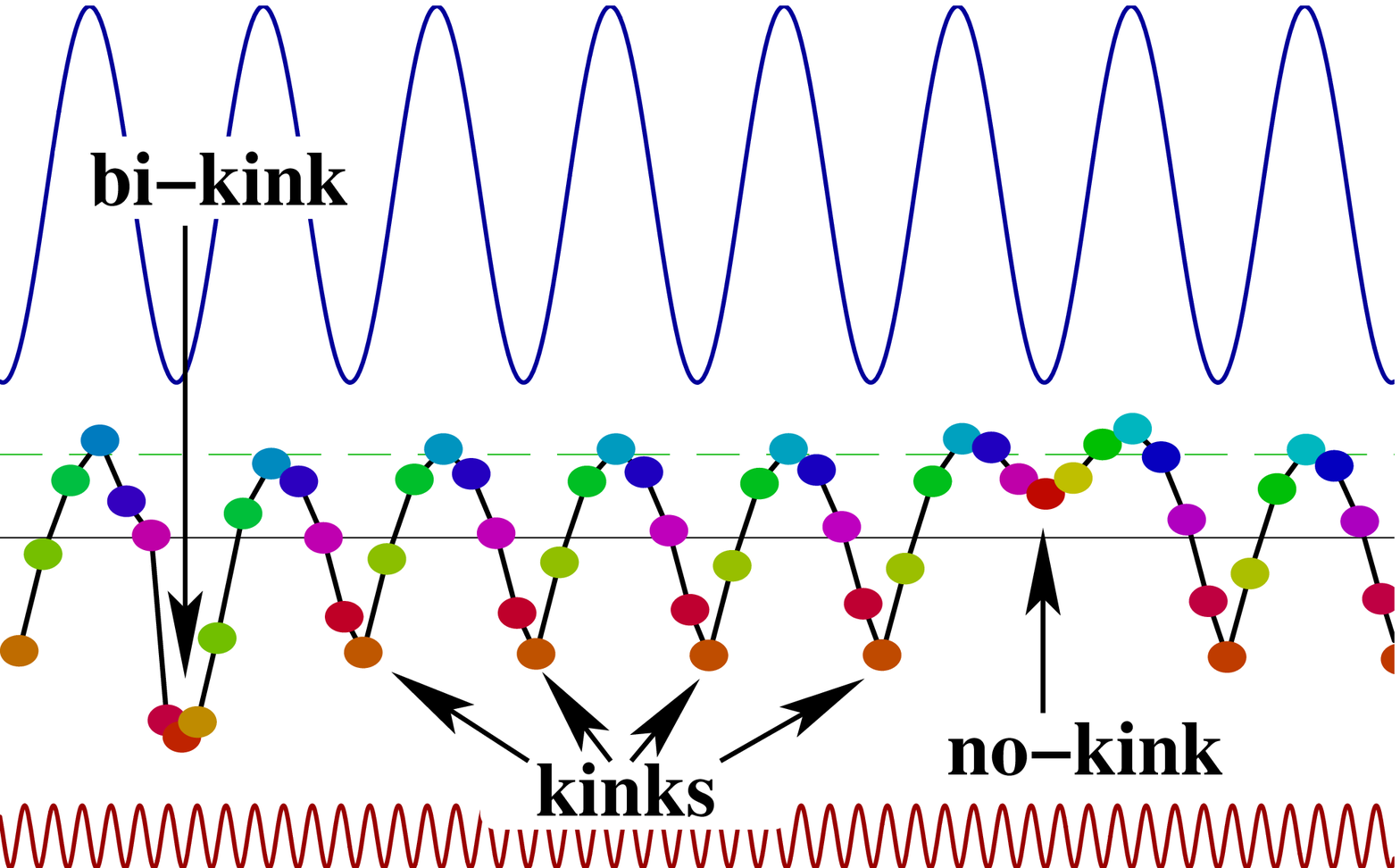,height=3cm,angle=0,clip=}
\qquad(b)
\epsfig{file=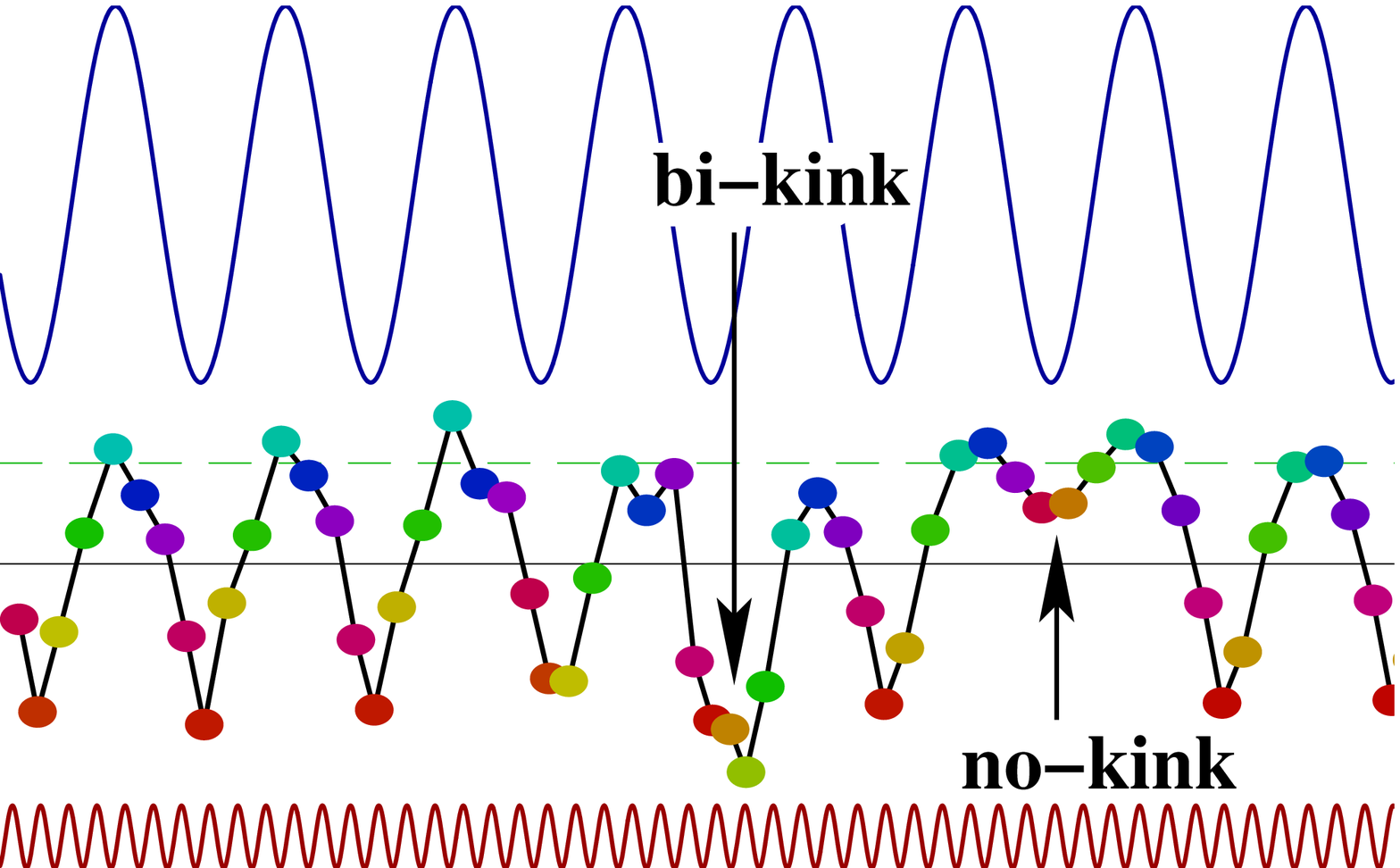,height=3cm,angle=0,clip=}
\hfill}
\vskip 3mm
\centerline{\hfill(c)
\epsfig{file=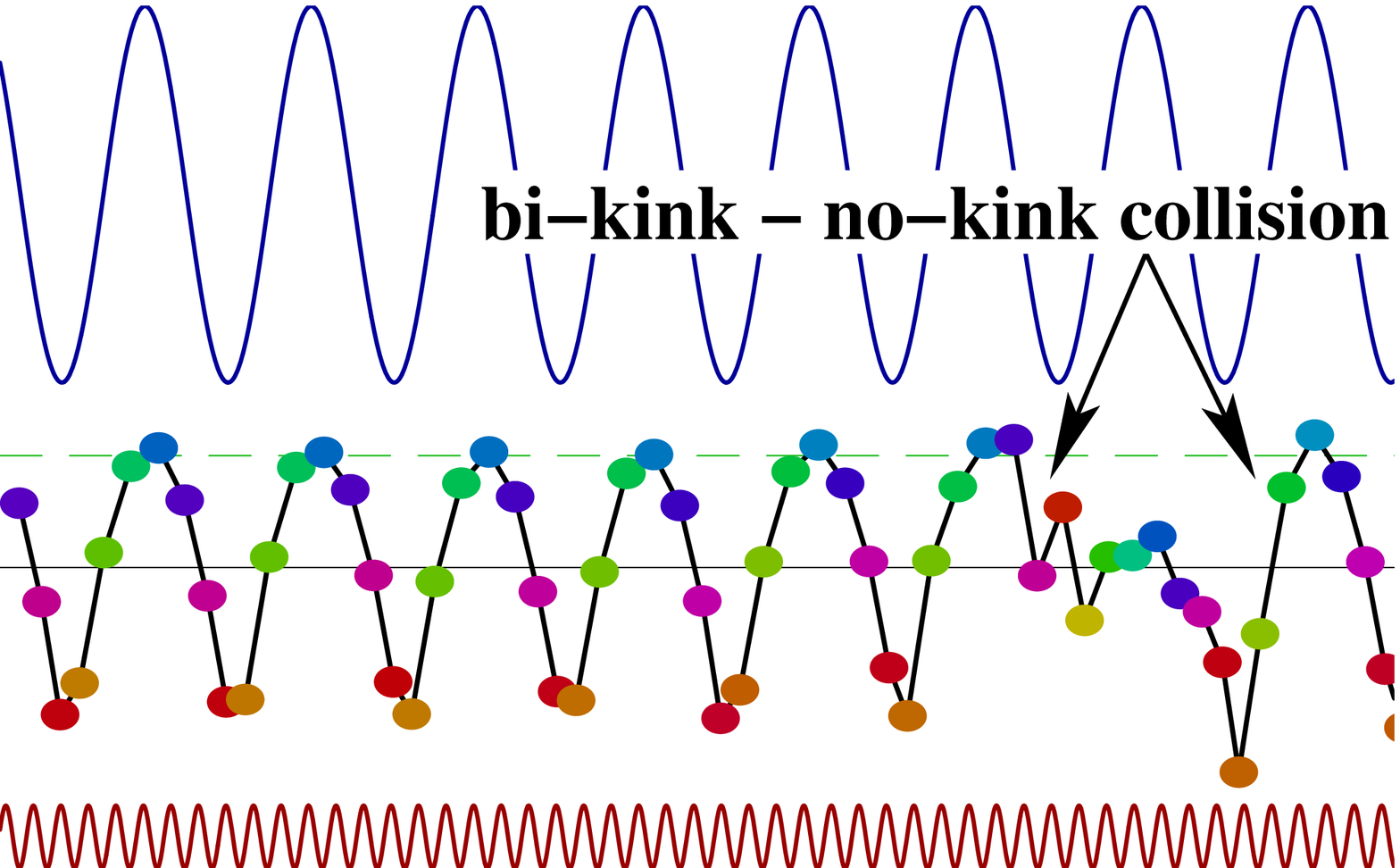,height=3cm,angle=0,clip=}
\qquad(d)
\epsfig{file=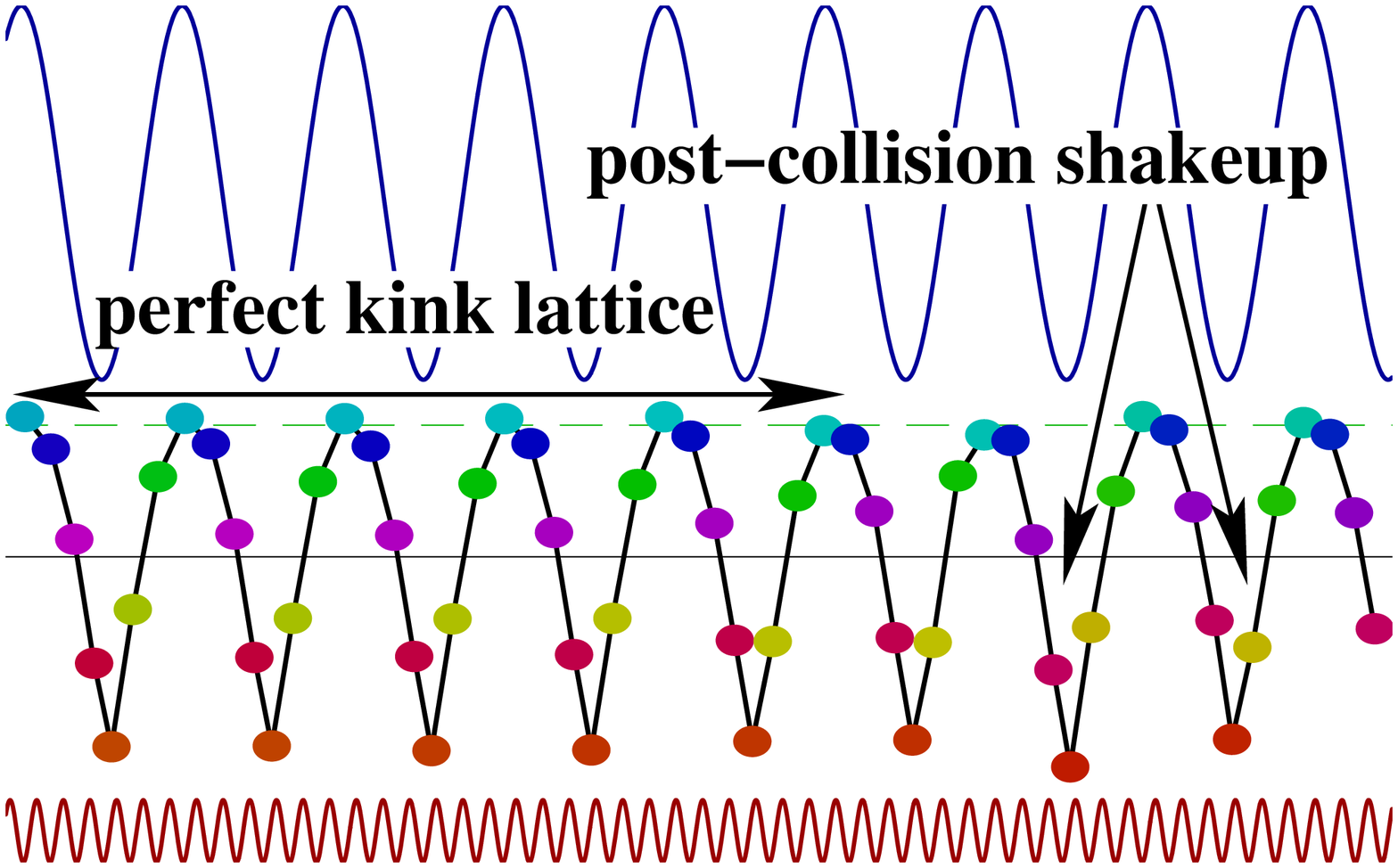,height=3cm,angle=0,clip=}
\hfill
}
\caption{\label{framesK5q1:fig}
(Color online)
Soft chain ($K$ = 5). Four snapshots of a 60-particles section of the
lubricant chain and $(+/-)$ substrates (lower/upper sinusoids), at
successive times separated by $14$ time units $(a_+\,m/ F_+)^{1/2}$.
The horizontal direction represents distance, the dots the particle
positions $x_i$.
Vertical displacements of dots measure the distance $x_i - x_{i-1}$ of a
given particle to its left neighbor: on this scale, the horizontal solid
and dashed lines indicate the average interparticle distance $a_0$, and the
$(+)$ lattice parameter $a_+$ respectively.
The snapshots refer to $r_+=(1+\pi^{1/4})/2$, $\theta=1$, $F_{\rm
ext}=0.08136$ (decreasing), and illustrate the crossing of the critical
line $F_c^{+\downarrow}$, with the recovery of the plateau state (see
Fig.~\ref{loop_hyst}) occurring through the disappearance of the last
bi-kink -- no-kink defect.
The other parameters are $\gamma=0.1$, and $v_{\rm ext}=0.1$.
This annihilation of a bi-kink against a stationary no-kink is best
illustrated by the online animation repinning\_theta1\_K5.gif, which spans
70 time units, starting 11 time units before frame (a) and ending 17 time
units after frame (d) of the present figure.
For this and all animations we select the reference frame where the $(-)$
substrate, and thus all pinned kinks, are stationary.
}
\end{figure}

For soft enough chains, individual kinks are visible and well distinct.
For example Fig.~\ref{framesK5q1:fig} (decreasing $F_{\rm ext}$)
illustrates the mechanism supporting deviations from the plateau for $K=5$,
the same value as Fig.~\ref{loop_hyst}.
A kink vanishes at a $(-)$ lattice site and joins a second kink to form a
mobile ``bi-kink''.  This extra density accumulation ``binds''
substantially less than a kink to the minima of the $(-)$ potential.
The external force $F_{\rm ext}$ acts on the bi-kink density lump and drags
it along to the right.
Contrary to the bi-kink, the site with a missing kink (``no-kink'') remains
pinned to the $(-)$ potential well, and is not dragged by the external
force $F_{\rm ext}$.
The moving bi-kink breaks the ``quantized'' motion by one single particle,
and is responsible for displacing the lubricant CM velocity a
little bit away from the exact $v_{\rm plateau}$.

The number of bi-kink -- no-kink pairs tends to increase rapidly with
increasing $F_{\rm ext}-F_c^{+\downarrow}$.
The force $F_c^{+\uparrow}$ necessary to nucleate the first bi-kink --
no-kink pair is sufficiently large to sustain an avalanche of more bi-kink
-- no-kink pairs after the first defect is nucleated.
Trains of bi-kinks cross the chain, producing essentially chaotic motions
of the single lubricant particles, provided that $F_{\rm ext}\gg
F_c^{+\downarrow}$.
When, starting from this dislodged, or depinned state, $F_{\rm ext}$ is
gradually reduced, bi-kink -- no-kink pairs annihilate, the number of these
pairs reducing steadily with time.
The discrete, integer nature of the defect-pair number originates the
(gently sloping) discrete downward staircase steps in the hysteresis loop,
generally similar to those shown in Fig.~\ref{loop_hyst} (for a different
$\theta$).
Since the discrete effect of the disappearance of a single defect-pair
becomes negligible in the infinite-size limit, the observed multi-step
structure appears to be merely a finite size artifact, and for all that we
can tell at present the infinite system should exhibit no staircase steps.
In the depinned state, so long as $F_{\rm ext}$ is strong enough, 
a bi-kink encounters a no-kink, interacts briefly, and then continues 
to travel.
When instead $F_{\rm ext}$ is reduced below $F_c^{+\downarrow}$, as in
Fig.~\ref{framesK5q1:fig}, the encounter of a bi-kink and a no-kink leads
to reciprocal annihilation.
The amplitude oscillation still visible (but quickly damped) at the right
end side of the last frame of Fig.~\ref{framesK5q1:fig} reflects the waves
dissipating the excess (``binding'') energy of the bi-kink -- no-kink pair,
in the process of recovering the perfect kink lattice.
When finally the kink lattice gets rid of the last defect pair, the perfect
plateau state is re-gained.

\begin{figure}
\centerline{\hfill(a)
\epsfig{file=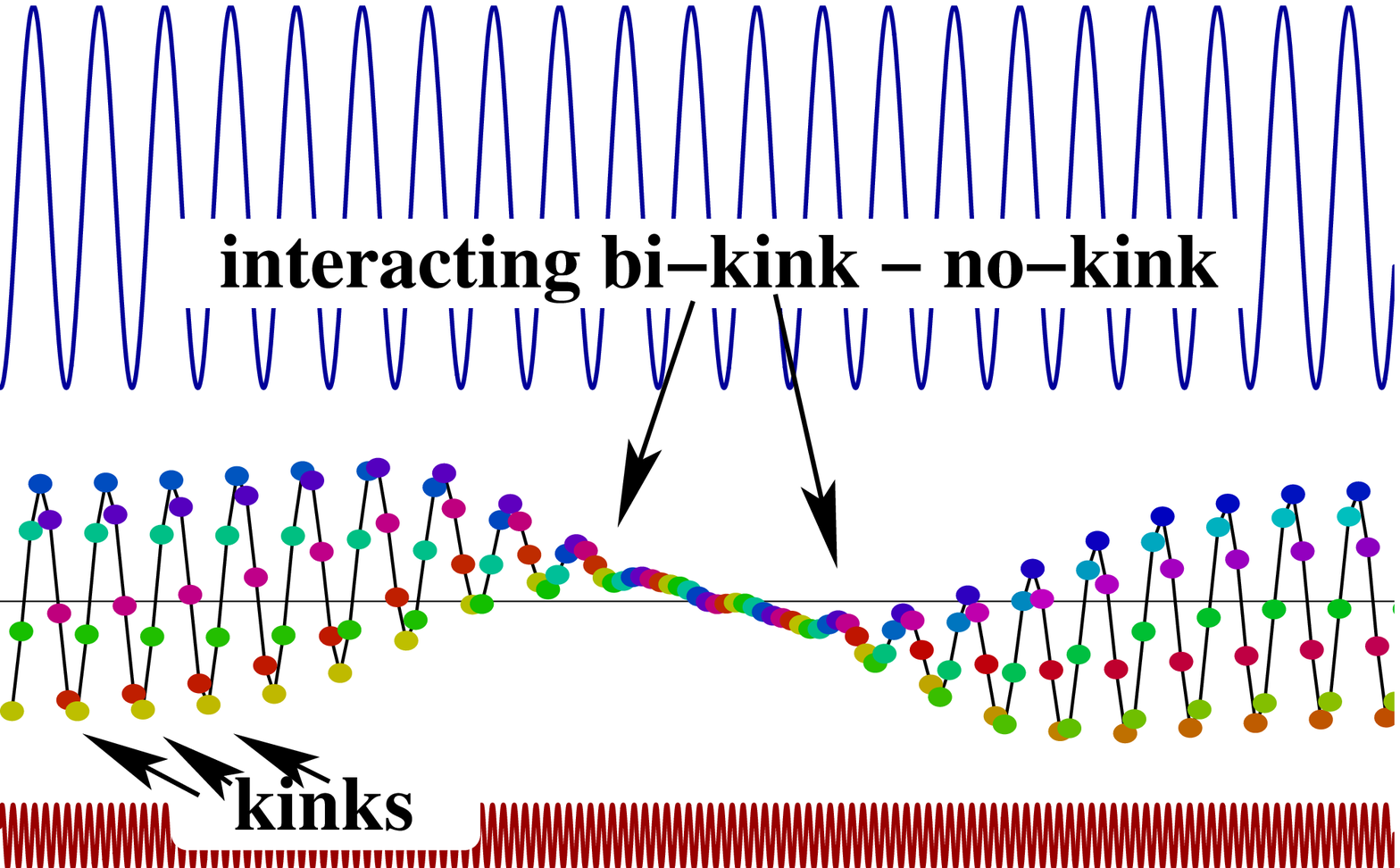,height=3cm,angle=0,clip=}
\qquad(b)
\epsfig{file=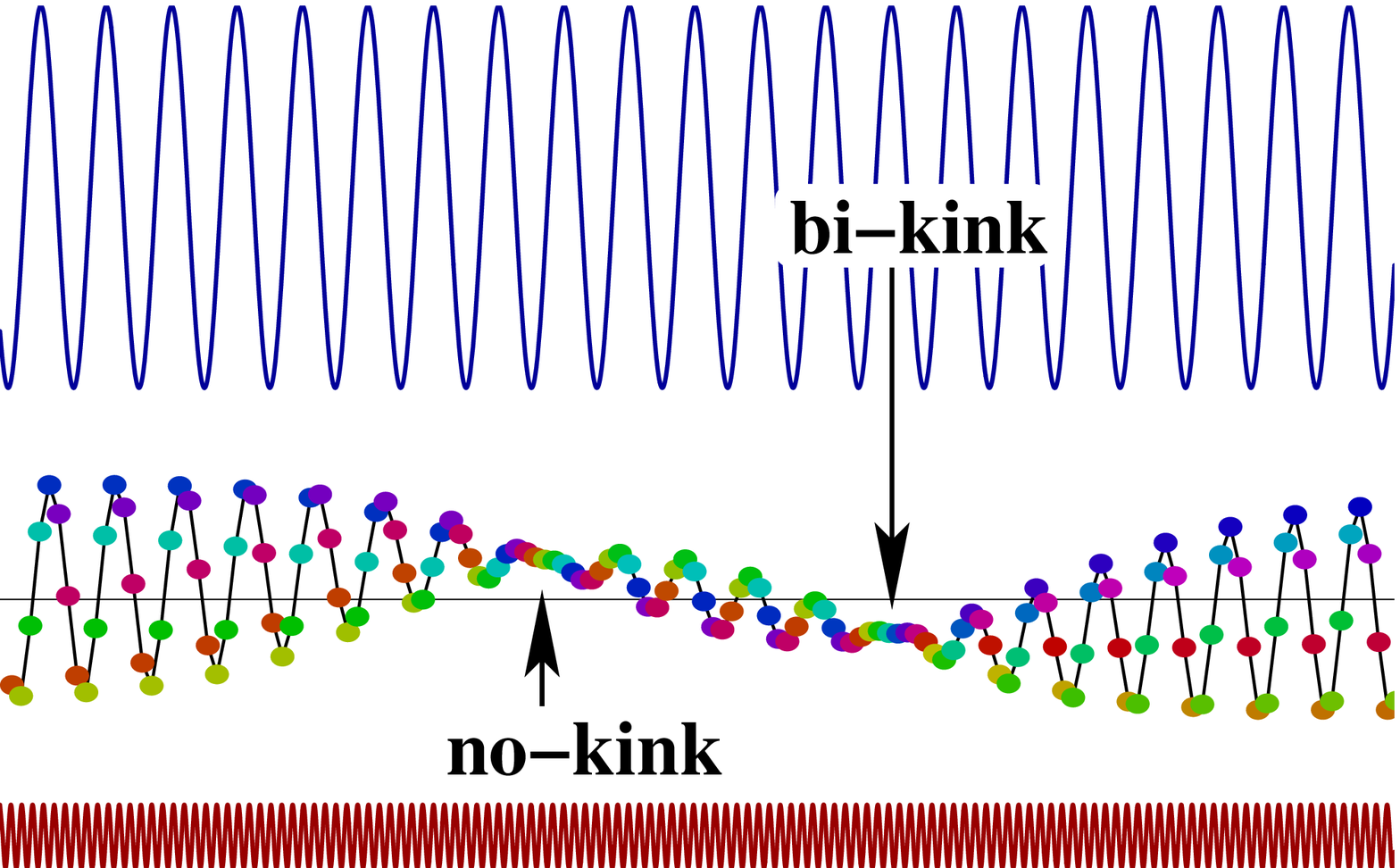,height=3cm,angle=0,clip=}
\hfill}
\vskip 3mm
\centerline{\hfill(c)
\epsfig{file=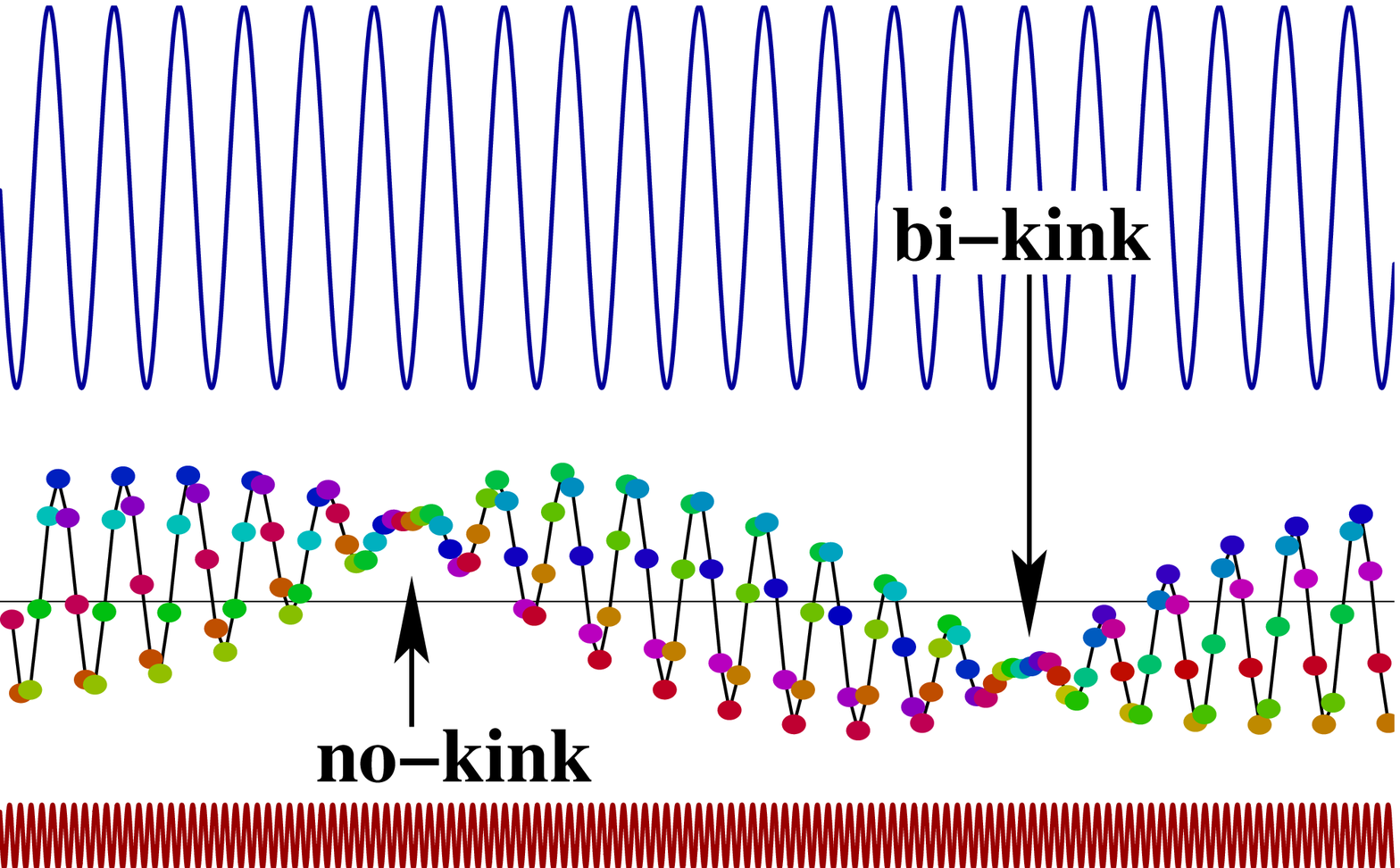,height=3cm,angle=0,clip=}
\qquad(d)
\epsfig{file=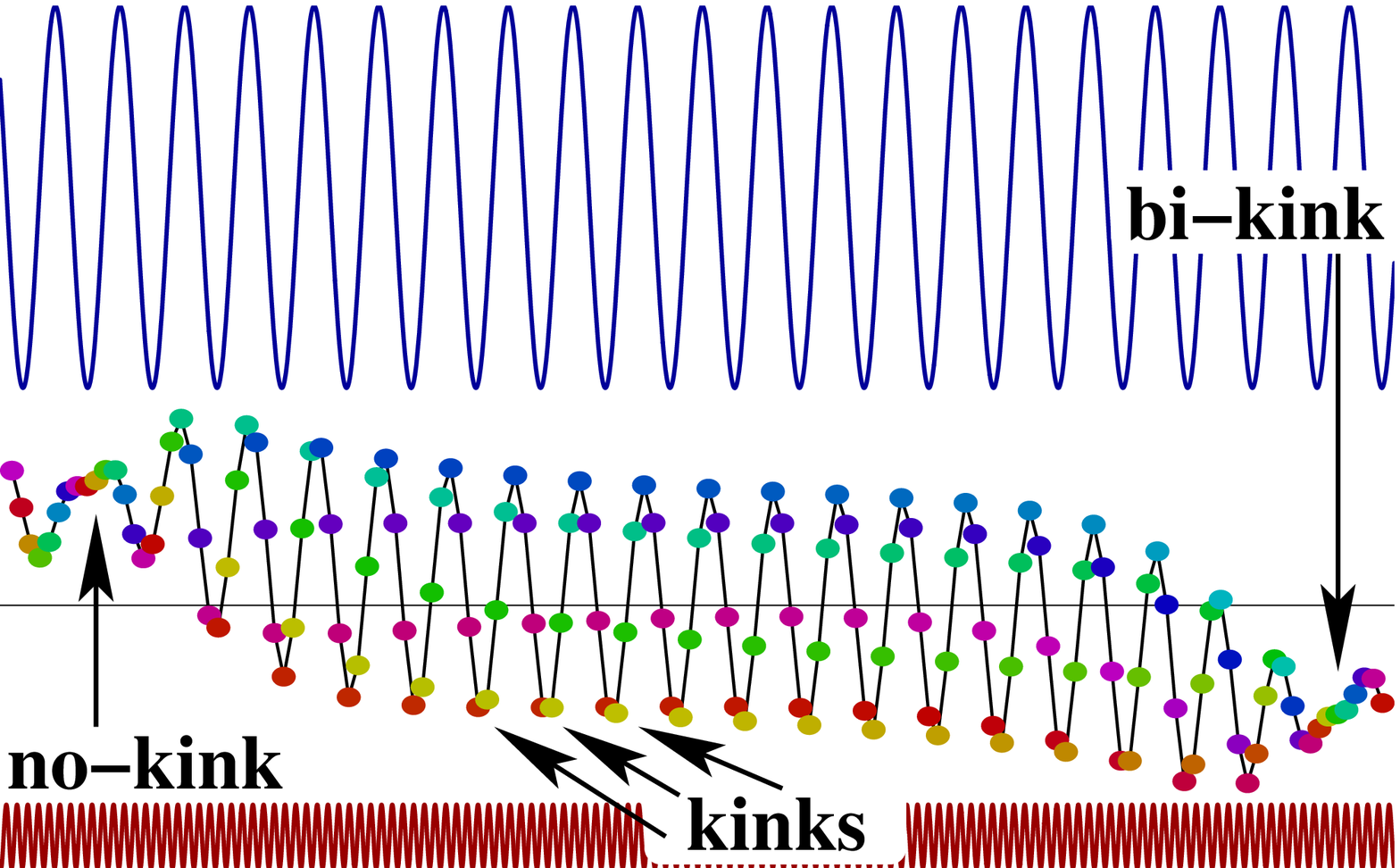,height=3cm,angle=0,clip=}
\hfill}
\caption{\label{framesq1:fig}
(Color online)
Stiff chain ($K$ = 50). Four successive snapshots of the substrates 
and lubricant chain, separated by time intervals of 9 model units.
All notations and parameters are the same as in Fig.~\ref{framesK5q1:fig},
except for $K=50$, $F_{\rm ext}=0.00685$ (decreasing), which falls in the
region immediately above the critical line $F_c^{+\downarrow}$, before the
recovery of the plateau state.
The complete collision of the right-traveling bi-kink and left-traveling
no-kink is best illustrated by the online animation
unpinned\_theta1\_K50.gif, which spans 50 time units, starting 16 time
units before frame (a) and ending 7 time units after frame (d).
}
\end{figure}

For a stiff enough chain, individual kinks become spatially broad, and will
for a fixed density extend over a size larger than the average inter-kink
distance $a_+/(r_+-1)$.
In this limit the kink lattice reduces to a weak sinusoidal deformation, of
amplitude $\propto K^{-1}$ superposed to the average interparticle density.
Despite this difference with the strong kink lattice of the soft-chain
case, the external-force--induced departure from the quantized velocity
plateau occurs here through a mechanism similar to that illustrated above
for the soft-spring case.
A chain slippage by one particle (i.e.\ a distance $a_0$) is promoted by a
bi-kink and a no-kink moving in opposite directions: when they collide, the
bi-kink -- no-kink pair takes the aspect of a broad locally flat region of
denser-than-average and less-dense-than-average lubricant in the otherwise
perfect pinned kink lattice.
As illustrated in Fig.~\ref{framesq1:fig}(a), a local flattening defect
forms in the soliton lattice, similar to the local amplitude suppression of
a dragged charge density wave (CDW) \cite{Gruener88,Inui88}.
This defect is characterized by a smooth ``charge'' separation, with the
denser region being driven to the right and the more rarefied region to the
left by the driving force, the external force acting like an electric field
on a CDW insulator.
These defects travel in opposite directions, as expected of a opposite
charges driven by an electric field.
The crucial difference with the soft-spring case (where as shown by
Fig.~\ref{framesK5q1:fig}, the no-kink defect remains pinned to the $(-)$
lattice) is that here both defects, the by-kink and the no-kink, are mobile
and dragged by the external force.
As the two defects move apart, a perfect soliton lattice re-forms in
between, Fig.~\ref{framesq1:fig}(d).
In time, a right-moving bi-kink encounters a left-moving no-kink: these
defects may again cross, or else they may bind and annihilate in pairs.
Annihilation occurs when $F_{\rm ext}$ is reduced below $F_c^{+\downarrow}$,
as in the soft-chain case of Fig.~\ref{framesK5q1:fig}.
When instead, $F_{\rm ext}>F_c^{+\downarrow}$ the pair separates again,
with the rightward ``positive'' and leftward ``negative'' flattenings
suffering some phase shift, but traveling on, as in
Fig.~\ref{framesq1:fig}.
As soon as all defects annihilate, the kink lattice is perfect, and the CM
velocity recovers $v_{\rm plateau}$ exactly.
If the defect pairs form at regular spatial separation within the chain
(with periodic boundary conditions) the corresponding moving pattern leads
to time-periodic fluctuations of the CM velocity; that can also be seen as
type-I intermittencies \cite{Berge84}.
Otherwise, when defect motion is chaotic, an irregular CM dynamics is
observed.
For indefinitely growing chain stiffness $K$, each defect pair flattening
region grows in size, eventually covering the entire finite-size
simulation, which becomes at that point a poor representation of the
infinite-size thermodynamical limit.

\begin{figure}
\centerline{\epsfig{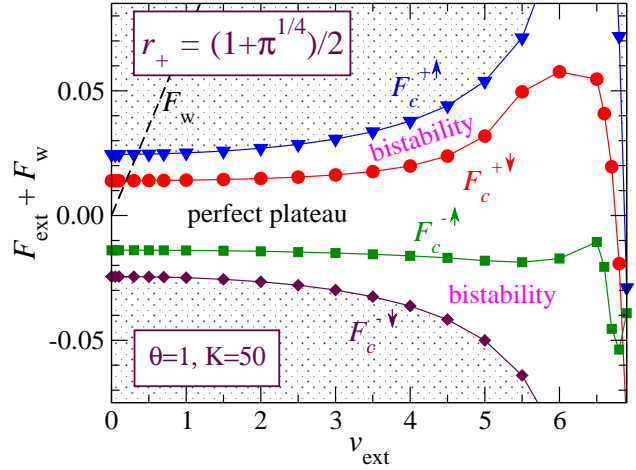}}
\caption{\label{FcVext:fig}
(Color online)
Driving-velocity dependency of the dynamical depinning and repinning forces
$F_c^{+\uparrow}$ $F_c^{+\downarrow}$, $F_c^{-\downarrow}$
$F_c^{-\uparrow}$ (shifted upward by the trivial $F_{\rm w}\propto v_{\rm
ext}$ contribution, Eq.~(\ref{Fw})).
$v_{\rm ext}$ is measured in model units of $(F_+a_+/m)^{1/2}$; the chain
is rather hard ($K=50$); $r_+=(1+\pi^{1/4})/2$, $\theta=1$ ($r_-\simeq
7.036$), and $\gamma=0.1$.
}
\end{figure}

Figure~\ref{FcVext:fig} draws the plateau boundaries relative to
$F_{\rm ext}$, for varied external driving $v_{\rm ext}$, for a rather
stiff chain ($K=50$).
As the friction-drag reference force $F_{\rm w}$ grows linearly with
$v_{\rm ext}$, and this introduces a trivial compensating trend $F_c^{\pm
\uparrow/\downarrow} \propto -F_{\rm w}$, it is convenient to remove the
appropriate linear drift by adding $F_{\rm w}$, Eq.~(\ref{Fw}), to the
critical forces.
The static limit $v_{\rm ext}=0$ is smooth, and this indicates a regime of
continuity from the static quasiperiodic 3-lengthscale model of
Ref.~\cite{Vanossi00} to the dynamical sliding.
Strikingly, the plateau robustness against the external perturbing force
$F_{\rm ext}$ and the widths of the hysteretical regions {\em benefit} of
increased driving speed.
For large $v_{\rm ext}\simeq 7$ the plateau destabilizes suddenly and
eventually disappears.

\subsection{Commensurate $\theta=5/4$}

\begin{figure}
\centerline{\hfill(a)
\epsfig{file=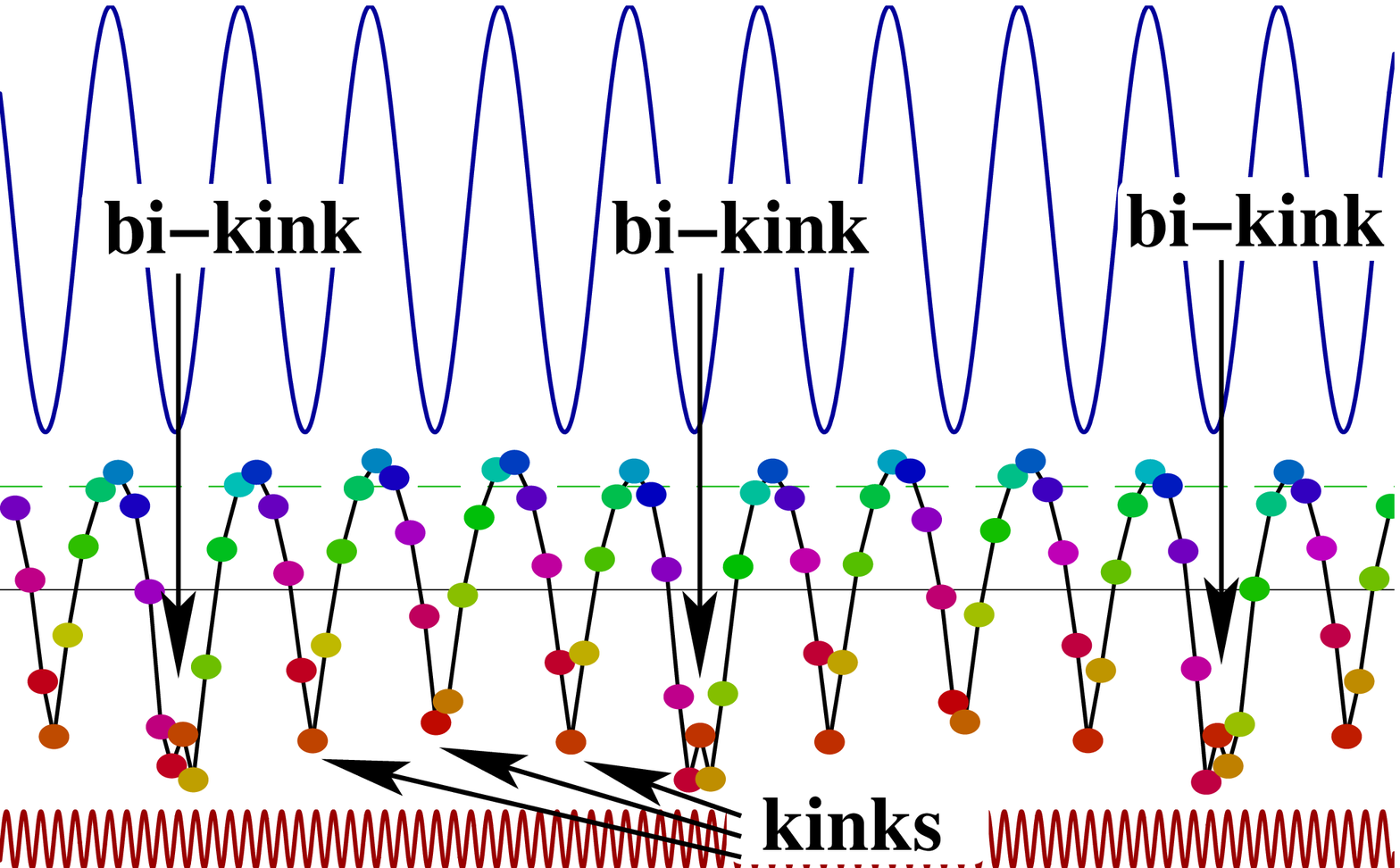,height=3cm,angle=0,clip=}
\qquad(b)
\epsfig{file=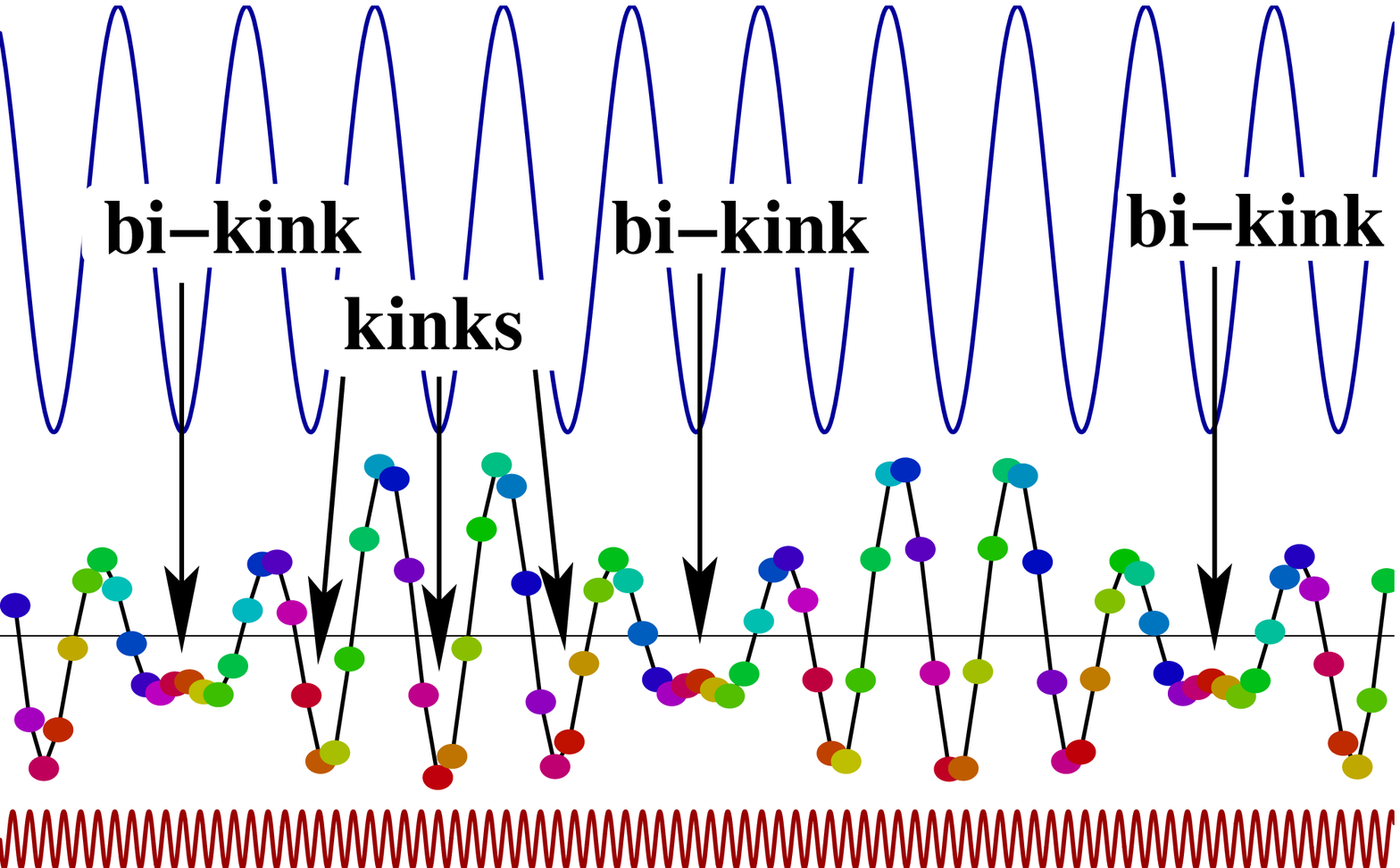,height=3cm,angle=0,clip=}
\hfill
}
\caption{\label{frame_plateau_q5f4:fig}
(Color online)
Typical plateau arrangements of the $\theta=5/4$ commensurate soft $K=5$
(a) and hard $K=50$ (b) chain:
a regular arrangement of bi-kinks (one every four kinks).
The conventions and all other parameters are the same as in
Fig.~\ref{framesK5q1:fig}, but for $F_{\rm ext}=-F_{\rm w}$.
}
\end{figure}

Having explored at length the $\theta=1$ commensurability, we now turn to
another kink lattice/slider system, still commensurate but with
$\theta=5/4$, a weaker commensurability than $\theta=1$.
At $\theta=5/4$, in the perfect-plateau state one kink out of four turns
into a bi-kink, as illustrated in Fig.~\ref{frame_plateau_q5f4:fig}.
(The bi-kinks of the present $\theta>1$ case would be replaced
by no-kinks for $\theta <1$).
The pre-existence of a regular array of such defects of the kink lattice
allows for a significantly different depinning mechanism, compared to the
totally commensurate $\theta=1$ case.
Defects of the kink lattice are already present prior to turning on the
external force $F_{\rm ext}$, which only sets them into motion, without a
need to create them.
For soft springs, Fig.~\ref{frame_plateau_q5f4:fig}(a), where the pinning
energy barrier of these defects is large, Fig.~\ref{criticalF} shows that
the critical forces needed to set the defects into motion in this
$\theta=5/4$ case are very similar to those for $\theta=1$.
For harder springs, defects increase in size and affect several neighboring
kinks now, as illustrated in Fig.~\ref{frame_plateau_q5f4:fig}(b).
These extended disturbances possess a much smaller pinning energy to the
$(-)$ potential.
As a consequence, the plateau state is now exceedingly weak, confined to an
extremely narrow force range around $-F_{\rm w}$, see Fig.~\ref{criticalF}.
The ordered arrangement of defects still warrants some amount of pinning,
but the width $F_c^{+\uparrow}-F_c^{-\downarrow}$ of the pinned region
decreases much faster than in the $\theta=1$ case as soon as the defect
size exceeds the typical inter-defect distance $a_+/(r_+-1)/\theta$, here
occurring for $K \simeq 10$.

\subsection{Incommensurate $\theta=(1+10^{1/2})/3$}

\begin{figure}
\centerline{\hfill(a)
\epsfig{file=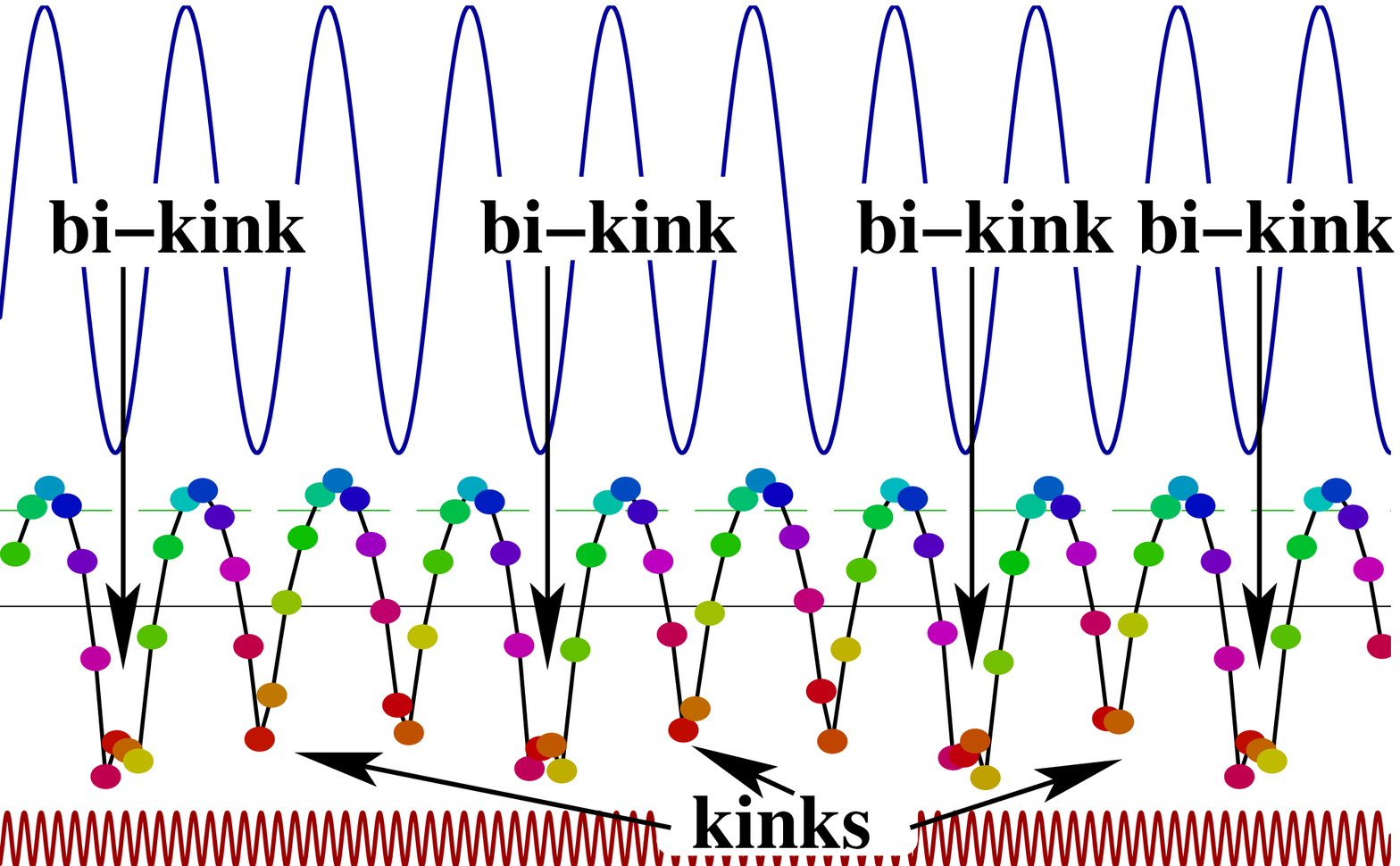,height=3cm,angle=0,clip=}
\qquad(b)
\epsfig{file=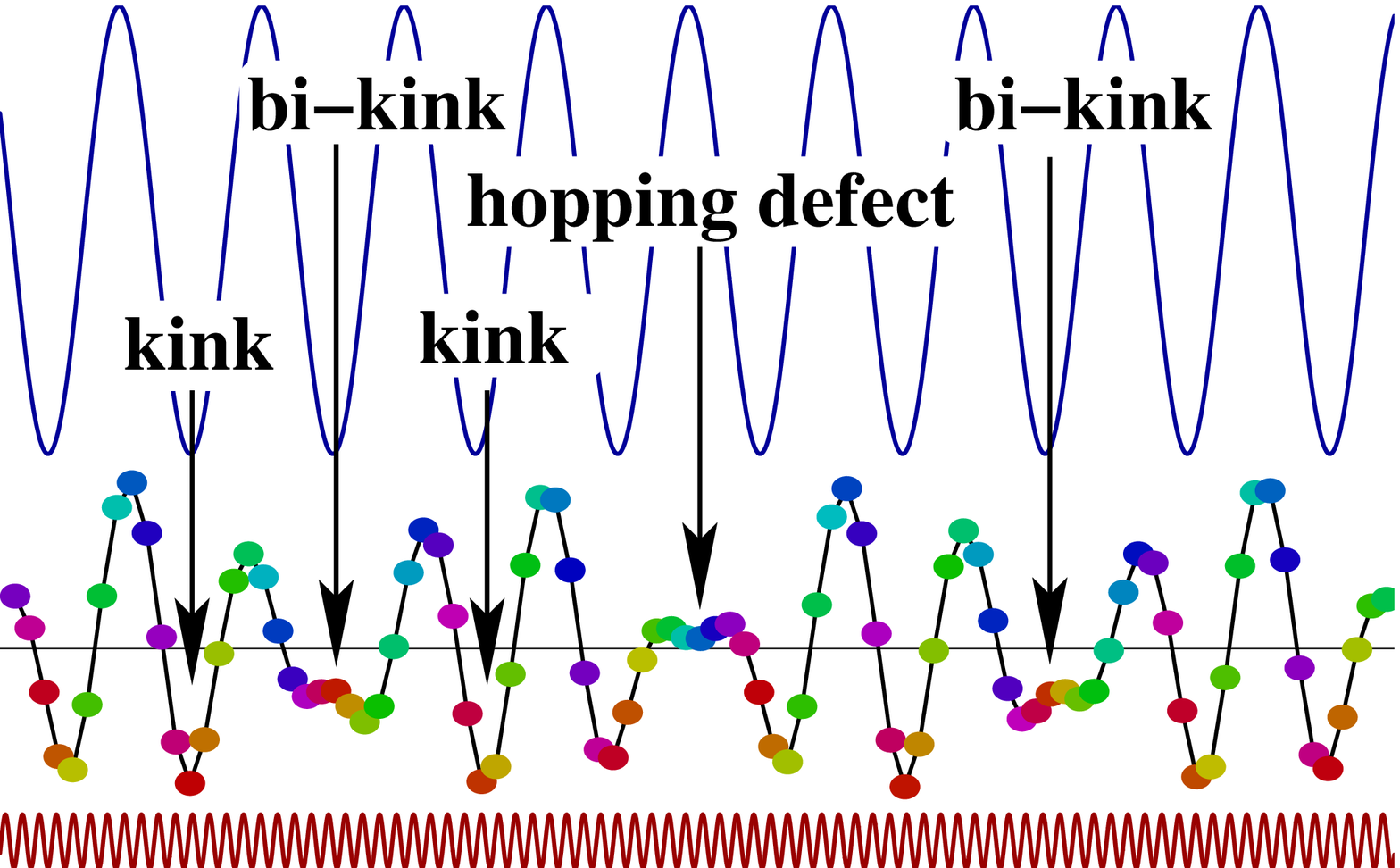,height=3cm,angle=0,clip=}
\hfill
}
\caption{\label{frame_plateau_q1plsq10f3:fig}
(Color online)
$\theta=(1+10^{1/2})/3$ incommensurate soft chain $K=5$ (a) and hard chain
$K=50$ (b): irregular alternation of kinks and bi-kinks.
Pinning is realized for the soft chain, while even with $F_{\rm
ext}=-F_{\rm w}$, so that $v_{\rm cm} \simeq v_{\rm plateau}$, the
$K=50$ hard chain is unpinned, with the defects slowly drifting along.
The conventions and all other parameters are the same as in
Fig.~\ref{framesK5q1:fig}.
The online file unpinned\_incommensurate.gif provides an animation of the
situation of snapshot (b).  A second multimedia file,
unpinned\_incommensurate\_drifting.gif, shows the defects drifting under
the effect of a deviation of $F_{\rm ext}$ by $10^{-3}$ force units in
excess of $-F_{\rm w}$.
}
\end{figure}

Finally, at irrational $\theta=(1+10^{1/2})/3$, some kinks are replaced by
bi-kinks, but the incommensuracy of the coverage leads to their irregular
arrangement, as illustrated in Fig.~\ref{frame_plateau_q1plsq10f3:fig}.
For a sufficiently soft chain, (represented by $K=5$ in
Fig.~\ref{frame_plateau_q1plsq10f3:fig}(a)), the irregular distribution of
single kinks and bi-kinks remains statically pinned to the minima of the
$(-)$ substrate, with a finite barrier to overcome for a bi-kink to migrate
to the next minimum.
This barrier guarantees the existence and robustness of the CM quantized
velocity plateau (with a first-order hysteretical boundary) in the present
incommensurate case, pretty much like for the commensurate cases.
This energy barrier protects the plateau against the movement of bi-kinks
until $K<K_{\rm Aubry}^{\rm dyn}\simeq 24$.
In contrast, for a harder chain ($K>K_{\rm Aubry}^{\rm dyn}$), illustrated
by $K=50$ in Fig.~\ref{frame_plateau_q1plsq10f3:fig}(b), the irregular
distribution of single kinks and bi-kinks drifts through the chain at a
speed approximately proportional to $F_{\rm ext}+F_{\rm w}$, with no sign
of any pinned plateau: this indicates that the energy barrier is here
entirely removed by the irregular bi-kink configuration produced by
incommensuracy.
The kink-kink repulsion makes the bi-kinks increasingly extended objects as
$K$ increases, until they become so broad that crossing the maxima of the
$(-)$ potential costs negligible energy: the bi-kink in the central region
of Fig.~\ref{frame_plateau_q1plsq10f3:fig}(b) exemplifies precisely one
such slow hopping process.
The transition between the soft-chain dynamically pinned regime and the
stiff-chain fully unpinned state is analogous to the Aubry transition
observed in the static situation described by the FK model.
The kinks of the dynamical model play the role of the particles of the
static model.

\subsection{Hysteresis when cycling other parameters}

By analogy to the single-chain FK model, cycling the external force $F_{\rm
ext}$ is conceptually the most natural way to abandon and recover, often
hysteretically, the dynamical plateau.
However, in practice, the experimental realization of a uniform force
acting equally on each lubricant particle in flight is not trivial.
On the other hand, the plateau can be abandoned and recovered, even when
different parameters are cycled.
Within the present model, the reason is that the dissipation $\gamma$-term
has itself the effect of diverting the CM velocity away from $v_{\rm
plateau}$.
In a concrete laboratory configuration moreover, beside dissipative
effects, other interactions too will tend to push the lubricant slide at
speeds other than $v_{\rm plateau}$.
As an example, defects and grain boundaries will tend to pin statically the
lubricant to either substrate \cite{Cesaratto07}.
These other ``external'' forces compete with the tendency to dynamical
pinning: the latter tuned by other parameters, namely, in the language of
our model, $K$ $F_+$ $F_-$ and $v_{\rm ext}$.
Thus in a practical straightforward experiment, cycling quantities such as
the the sliding speed, or the load applied to the sliders should lead to
leaving/recovering the plateau dynamics, with hysteretic cycles similar to
those exemplified by Fig.~\ref{loop_hyst}.

\begin{figure}
\centerline{\epsfig{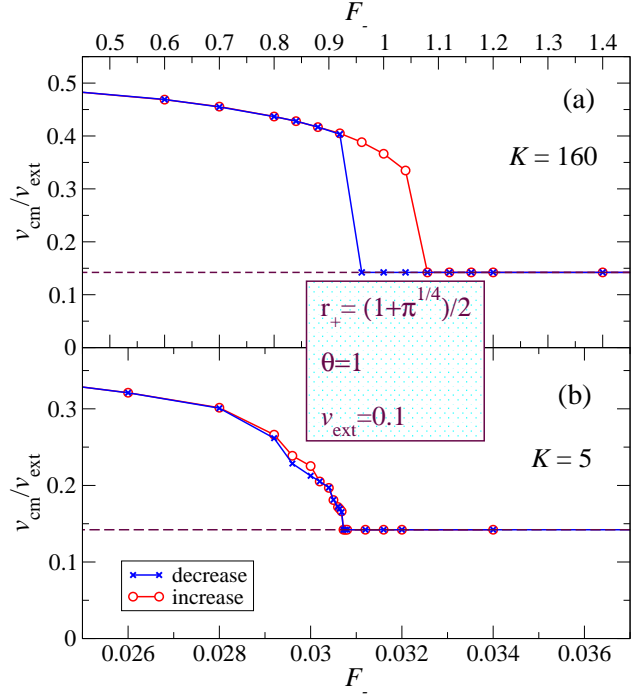}}
\caption{\label{loop_FMhyst}
(Color online)
Hysteresis loops found as the corrugation of the $(-)$ substrate $F_-$ is
cycled down from (crosses) and back up to (circles) its value $F_+$ used in
all other calculations.
(a) At $K=160$, near the plateau edge of Fig.~\ref{velcm_model}, it takes a
small decrease in $F_-$ to leave the plateau, while (b) when the plateau is
very robust ($K=5$), nonhysteretic depinning is observed for a corrugation
amplitude $F_-$ far below unity.
Simulations for $r_+=(1+\pi^{1/4})/2$, $\theta=1$ ($r_-\simeq 7.036$),
$\gamma=0.1$, $v_{\rm ext}=0.1$.
}
\end{figure}

To illustrate this point within our model, Fig.~\ref{loop_FMhyst} depicts a
first example of such a hysteretic cycle, where the load applied to the
sliders, proportional to the upper slider corrugation $F_-$, is cycled.
The plateau is abandoned hysteretically when $F_-$ is decreased below
critical values which depend strongly on the robustness of the pinned
state, which is, in turn, a function of $K$ and other model parameters.

\begin{figure}
\centerline{\epsfig{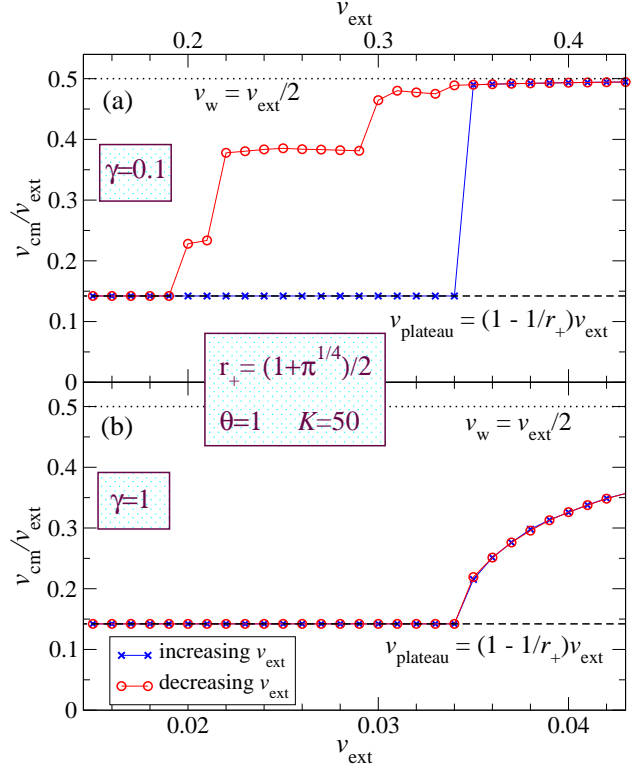}}
\caption{\label{Vloop_hyst}
(Color online)
(a) Hysteresis loop at the plateau edge in the $v_{\rm cm}-v_{\rm ext}$
characteristics for a confined chain of length ratios
$r_+=(1+\pi^{1/4})/2$, $\theta=1$ ($r_-=7.036$).
Adiabatic increase and decrease of $v_{\rm ext}$
are denoted by crosses and circles, respectively.
Here $\gamma=0.1\,(F_+m/a_+)^{1/2}$ and $F_{\rm ext}=0$, which 
corresponds to the dashed path of Fig.~\ref{FcVext:fig}.
(b) No hysteresis is observed in the overdamped regime
($\gamma=1.0\,(F_+m/a_+)^{1/2}$) along the same path.
}
\end{figure}

Along a similar scheme, the perfectly legitimate interpretation of
Fig.~\ref{FcVext:fig} as a phase diagram suggests that the first-order line
separating the free-sliding regime from the perfect plateau could be
crossed by cycling $v_{\rm ext}$ rather than $F_{\rm ext}$.
This cycle corresponds to tracking up and down the $F_{\rm ext}=0$ dashed
path drawn in Fig.~\ref{FcVext:fig}.
The resulting loop, shown in Fig.~\ref{Vloop_hyst}(a), depicts the expected
bistability: $v_{\rm ext}$ is cycled up and down, and the perfect plateau
is abandoned at much larger speed than where it is recovered.
At large speed $F_{\rm w}$ increases, the dissipative term dominates and
makes the lubricant speed approach $v_{\rm w}$.

The depinning transition may also occur continuously, when the transition
line is crossed beyond the tricritical point, i.e.\ for $K>K_*$, in the
strongly dissipative region, where the the viscous damping rate $\gamma/m$
is much larger than the vibrational frequencies, decreasing proportionally
to $K^{-1}$, of the soft kink lattice around the minima of the $(-)$
potential.
In this regime the dynamical depinning is apparently second order.
In this overdamped regime, shown for example in Fig.~\ref{Vloop_hyst}(b),
the forward and backward trajectories become indistinguishable, and
hysteresis disappears.
In this strongly dissipative regime, we find, instead of the hysteretic
jumps, a nonlinear dependency of $v_{\rm cm}$ versus the model parameters
(here $v_{\rm ext}$, but cycling $F_{\rm ext}$, $F_-$, or $K$ would lead to
perfectly analogous results), without any bistability phenomena.

\section{Discussion and conclusions}
%
We have shown that starting from the quantized sliding plateau state,
previously found for a simple tribological model of a confined layer, the
sliding dynamics of the lubricant layer exhibits a large hysteresis when an
additional external driving force $F_{\rm ext}$ trying to push $v_{\rm cm}$
away from its quantized value is cycled.
In analogy to depinning in ordinary static friction \cite{Braun97}, the
hysteretic dynamical behavior depends strongly on whether the system
degrees of freedom have sufficient inertia (underdamped regime) or if, on
the contrary, inertia is negligible (overdamped regime).
Hysteretic versus continuous depinning occurs depending on whether the
unpinning transition is crossed below or above a tricritical point where
hysteresis close, and which marks the separation between the underdamped
and the overdamped dynamics.

Hysteresis arises due to the great robustness of the quantized dynamics,
setting a large critical threshold $F_c^{+\uparrow}$ to the formation of
mobile defects (initially depinned bi-kinks or no-kinks).
Once at least one of these defects forms, an avalanche process leads to a
discontinuous jump to a free or quasi-free sliding regime.
Starting from the unpinned states, the plateau recovers only at a much
smaller threshold $F_c^{+\downarrow}$, representing the minimum driving
force needed to sustain the motion of pre-existing mobile defects.

Nontrivial differences with static friction occur. The first is that the
dynamical pinning hysteresis cycle may be larger in situations where
pinning itself could be intuitively considered more fragile, e.g., for
larger external velocity.
Another feature (presently under investigation, not discussed above) is
that the {\em sudden} application of an external force can sometimes leave
$v_{\rm cm}$ locked to the quantized value, even if the applied force is
larger than the dynamic depinning threshold $F_c^{+\uparrow}$ obtained
instead through the adiabatic procedure sketched above.
Once again, this is different from static depinning, usually requiring
smaller force (than the static friction $F_s$) if applied suddenly
\cite{Braunbook}.

The present study concentrates on zero temperature.
At finite temperature, the energy barrier to the formation of defects such
as bi-kinks and for defects ``hopping'' to neighboring pinning sites can be
traversed by means of random thermal excitations.
This means that at sufficiently low temperature the dynamical pinning
should not change much.
Even the hysteresis should remain, provided that parameters such as $F_{\rm
ext}$ are cycled much faster than the characteristic thermal relaxation
times.
Thermal effects are currently under closer investigation.

\ack

This research was partially supported by PRRIITT (Regione Emilia
Romagna), Net-Lab ``Surfaces \& Coatings for Advanced Mechanics
and Nanomechanics'' (SUP\&RMAN) and by  PRIN Cofin 2006022847,
as well as by INFM/CNR ``Iniziativa trasversale calcolo parallelo''.

\section*{References}


\begin{thebibliography}{10}

\bibitem{Kapitaniak99}
{ T.\ Kapitaniak and J.\ Wojewoda, \textit{Attractors of quasiperiodically
  forced systems} (World Scientific, Singapore, 1993)}.

\bibitem{Perssonbook}
{ B.\ N.\ J.\ Persson, \textit{Sliding Friction: Physical Principles and
  Applications (NanoScience and Technology)} (Springer-Verlag, Berlin 1998)}.

\bibitem{Rubinstein04}
{ S.\ M.\ Rubinstein, G Cohen, and J.\ Fineberg, Nature {\bf 430}, 1005
  (2004)}.

\bibitem{Vanossi_review}
{ See, e.g., A.\ Vanossi and O.\ M.\ Braun, ``Simulation of nanofriction
  through driven simplified models'', to appear in \textit{Advances in contact
  mechanics}, edited by R.\ Buzio and U.\ Valbusa (Research Signpost, Kerala,
  India), and references therein}.

\bibitem{Braunbook}
{ O.\ M.\ Braun and Yu.\ S.\ Kivshar, \textit{The Frenkel-Kontorova Model:
  Concepts, Methods, and Applications} (Springer-Verlag, Berlin, 2004)}.

\bibitem{Rozman96}
{ M.\ G.\ Rozman, M.\ Urbakh, and J.\ Klafter, Phys.\ Rev. Lett.\ {\bf 77}, 683
  (1996); Europhys.\ Lett. {\bf 39}, 183 (1997)}.

\bibitem{Zaloj98}
{ V.\ Zaloj, M.\ Urbakh, and J.\ Klafter, Phys.\ Rev. Lett.\ {\bf 81}, 1227
  (1998)}.

\bibitem{Dienwiebel04}
{ M.\ Dienwiebel, G.\ S.\ Verhoeven, N.\ Pradeep, J.\ W.\ M.\ Frenken, J.\ A.\
  Heimberg, and H.\ W.\ Zandbergen, Phys.\ Rev.\ Lett.\ {\bf 92}, 126101
  (2004)}.

\bibitem{Salmeron}
{ J.\ Y.\ Park, D.\ F.\ Ogletree, M.\ Salmeron, R.\ A.\ Ribeiro, P.\ C.\
  Canfield, C.\ J.\ Jenks, and P.\ A.\ Thiel, Science {\bf 309}, 1354 (2005)}.

\bibitem{Braun05}
{ O.\ M.\ Braun, A.\ Vanossi, and E.\ Tosatti, Phys.\ Rev.\ Lett.\ {\bf 95},
  026102 (2005)}.

\bibitem{Vanossi06}
{ A.\ Vanossi, N.\ Manini, G.\ Divitini, G.\ E.\ Santoro, and E.\ Tosatti,
  Phys.\ Rev. Lett.\ {\bf 97}, 056101 (2006)}.

\bibitem{Santoro06}
{ G.\ E.\ Santoro, A.\ Vanossi, N.\ Manini, G.\ Divitini, and E.\ Tosatti,
  Surf.\ Sci. {\bf 600}, 2726 (2006)}.

\bibitem{Manini07extended}
{ N.\ Manini, M.\ Cesaratto, G.\ E.\ Santoro, E.\ Tosatti, and A.\ Vanossi, J.\
  Phys.: Condens.\ Matter {\bf 19}, 305016 (2007)}.

\bibitem{Vanossi07PRL}
{ A.\ Vanossi, N.\ Manini, F.\ Caruso, G.\ E.\ Santoro, and E.\ Tosatti,
  submitted to Phys.\ Rev. Lett. (2007)}.

\bibitem{Braun97}
{ O.\ M.\ Braun, A.\ R.\ Bishop, and J.\ R\"{o}der, Phys.\ Rev.\ Lett.\ {\bf
  79}, 3692 (1997)}.

\bibitem{Ariyasu87}
{ J.\ C.\ Ariyasu and A.\ R.\ Bishop, Phys.\ Rev.\ B {\bf 35}, 3207 (1987)}.

\bibitem{Vanossi07Hyst}
{ A.\ Vanossi, G.\ E.\ Santoro, N.\ Manini, M.\ Cesaratto, and E.\ Tosatti, in
  press Surf.\ Sci. (2007); cond-mat/0609117}.

\bibitem{Gruener88}
{ G.\ Gr\"uner, Rev.\ Mod.\ Phys.\ {\bf 60}, 1129 (1988)}.

\bibitem{Inui88}
{ M.\ Inui, R.\ P.\ Hall, S.\ Doniach, and A.\ Zettl, Phys.\ Rev.\ B\ {\bf 38},
  13047 (1988)}.

\bibitem{Berge84}
{ P.\ Berg\'e, Y.\ Pomeau, and C.\ Vidal, \textit{Order within Chaos} (Hermann
  and Wiley, Paris, 1984)}.

\bibitem{Vanossi00}
{ A.\ Vanossi, J.\ R\"{o}der, A.\ R.\ Bishop, and V.\ Bortolani, Phys.\ Rev. E
  {\bf 63}, 017203 (2000)}.

\bibitem{Cesaratto07}
{ M.\ Cesaratto, N.\ Manini, A.\ Vanossi, E.\ Tosatti, and G.\ E.\ Santoro, in
  press Surf.\ Sci. (2007); cond-mat/0609116}.

\end{thebibliography}

\end{document}